\documentclass[12pt]{article}
\usepackage[centertags]{amsmath}
\usepackage{amssymb,amsfonts}
\usepackage[comma,super]{natbib}
\usepackage{latexsym}
\newcommand{\be}{\begin{equation}} \newcommand{\ee}{\end{equation}}
\newcommand{\bea}{\begin{eqnarray}} \newcommand{\eea}{\end{eqnarray}}
\newcommand{\bse}{\begin{subequations}} \newcommand{\ese}{\end{subequations}}

\begin{document}
\begin{center}
{\Large \bf Charged anisotropic matter with linear equation of state} \\
\vspace{1.5cm} {\bf S. Thirukkanesh$^\dag$ and S. D. Maharaj}\\
Astrophysics and Cosmology Research Unit,\\
School of Mathematical Sciences,\\
University of KwaZulu-Natal,\\
Private Bag X54001,\\
Durban 4000,\\
South Africa.\\
\vspace{1.5cm} {\bf Abstract}\\
\end{center}
We consider the general situation of a compact relativistic body
with anisotropic pressures in the presence of the electromagnetic
field. The equation of state for the matter distribution is linear
and may be applied to strange stars with quark matter. Three classes
of new exact solutions are found to the Einstein-Maxwell system.
This is achieved by specifying a particular form for one of the
gravitational potentials and the electric field intensity.  We can
regain anisotropic and isotropic models from our general class of
solution. A physical analysis indicates that the charged solutions
describe realistic compact spheres with anisotropic matter
distribution. The equation of state is consistent with dark energy
stars and charged quark matter distributions. The masses and central
densities correspond to realistic stellar objects in the general
case when anisotropy and charge are present.
 \vspace{5cm}
~\\
$^\dag$Permanent address: Department of Mathematics, Eastern
University, Sri Lanka, Chenkalady, Sri Lanka.

\section{Introduction}
Since the pioneering paper by Bowers and Liang [1] there have been
extensive investigations in the study of anisotropic relativistic
matter distributions in general relativity to include the effects of
spacetime curvature. The anisotropic interior spacetime matches to
the Schwarzschild exterior model. The early work of Ruderman [2]
showed that nuclear matter may be anisotropic in density ranges of
$10^{15}$ gcm$^{-3}$ where nuclear interactions need to be treated
relativistically. Note that conventional celestial bodies are not
composed purely of perfect fluids so that radial pressures are
different from tangential pressures. Anisotropy can be introduced by
the existence of a solid stellar core or by the presence of a type
3A superfluid as indicated by Kippenhahn and Weigert [3]. Different
kinds of phase transitions  (Sokolov [4]) or pion condensation
(Sawyer [5]) can generate anisotropy. Binney and Tremaine [6] have
considered anisotropies in spherical galaxies in the context of
Newtonian gravitational theory. Herrera and Santos [7] studied the
effects of slow rotation in stars and Letelier [8] analysed the
mixture of two gases, such as ionized hydrogen and electrons, in a
framework of a relativistic anisotropic fluid. Weber [9] showed that
strong magnetic fields serve as a vehicle for generating anisotropic
pressures inside a compact sphere. Some recent anisotropic models
for compact self-gravitating objects with strange matter include the
results of  Lobo [10] and Sharma and Maharaj [11] with a barotropic
equation of state. Therefore the study of anisotropic fluid spheres
in static spherically symmetric spacetimes is important in
relativistic astrophysics.

In recent years there have been several investigations of the
Einstein-Maxwell system of equations for static spherically
symmetric gravitational fields usually with isotropic pressures to
include the effects of the electromagnetic field. The interior
spacetime must match at the boundary to the Reissner-Nordstrom
exterior model. The models generated can be used to describe charged
relativistic bodies in strong gravitational fields such as neutron
stars. Many exact solutions have been given by Ivanov [12] and
Thirukkanesh and Maharaj [13] which satisfy the conditions for a
physically acceptable charged relativistic sphere. Charged
spheroidal stars have been studied extensively by Komathiraj and
Maharaj [14], Sharma $et~ al$ [15], Patel and Koppor [16], Tikekar
and Singh [17] and Gupta and Kumar [18]. These charged spheroidal
models contain uncharged neutron stars in the relevant limit and are
consequently relevant in the description of dense astrophysical
objects. We point out the particular detailed studies of Sharma $et~
al$ [19] in cold compact objects,  Sharma and Mukherjee [20]
analysis of strange matter and binary pulsars, and Sharma and
Mukherjee [21] analysis of quark-diquark mixtures in equilibrium in
the presence of the electromagnetic field. Charged relativistic
matter is also relevant in modeling core-envelope stellar system as
shown in the treatments of Thomas $et~al$ [22], Tikekar and Thomas
[23] and Paul and Tikekar [24] in which the stellar core is an
isotropic fluid surrounded by a layer of anisotropic fluid.
Consequently the study of charged fluid spheres in static
spherically symmetric spacetimes is of significance in relativistic
astrophysics.

From the above motivation it is clear that both anisotropy and the
electromagnetic field are important in astrophysical processes.
However previous treatments have largely considered either
anisotropy or electromagnetic field separately. The intention of
this paper is to provide a general framework that admits the
possibility of tangential pressures with a nonvanishing electric
field intensity. We believe that this approach will allow for a
richer family of solutions to the Einstein-Maxwell field equations
and possibly provide a deeper insight into the behaviour of the
gravitational field. On physical grounds we impose a barotropic
equation of state which is linear, that relates the radial pressure
to the energy density and allows for the existence of strange
matter. Our general model will contain strange matter solutions
found previously. In this regard we mention the following recent
works on strange stars. Mak and Harko [25] and Komathiraj and
Maharaj [26] found analytical models in the MIT bag model (Witten
[27]) with a strange matter equation of state in the presence of an
electromagnetic field. Sharma and Maharaj  [11] generated a class of
exact solutions which can be applied to strange stars with quark
matter for neutral anisotropic matter. Lobo [10] found stable dark
energy stars which generalise the gravastar model governed by a dark
energy equation of state.

The objective of this treatment is to generate exact solutions to
the Einstein-Maxwell system, with linear equation of state, that may
be utilised to describe a charged anisotropic relativistic body. In
Section 2, we express the Einstein-Maxwell system as a new system of
differential equations using a coordinate transformation, and then
write the system in another form which is easier to analyse. Three
classes of new exact solutions to the Einstein-Maxwell system are
found in Section 3 in terms of simple elementary functions. We show
that particular uncharged anisotropic strange stars found in the
past are contained in our general family of solutions. In Section 4,
we show that the solutions are physically admissible and plot the
matter variables for particular parameter values. We generate values
for the mass and central density in Section 5 for charged and
uncharged matter. This analysis extends the treatment of Sharma and
Maharaj [11] to include charge, and confirms that the exact
solutions found are physically reasonable. Some concluding remarks
are made in Section 6.

\section{The field equations}
Our intention is to model the interior of a dense star. On physical
grounds it is necessary for the gravitational field to be static and
spherically symmetric. Consequently, we assume that the interior of
a spherically symmetric star is described by the line element
\begin{equation}
\label{eq:f1} ds^{2} = -e^{2\nu(r)} dt^{2} + e^{2\lambda(r)} dr^{2}
+
 r^{2}(d\theta^{2} + \sin^{2}{\theta} d\phi^{2})
\end{equation}
in  Schwarzschild coordinates $(x^{a}) = (t,r,\theta,\phi).$ We take
the energy momentum tensor for an anisotropic charged imperfect
fluid sphere to be of the form
\begin{equation}
\label{eq:f2} T_{ij}=\mbox{diag}(-\rho -\frac{1}{2}E^2, p_r-
\frac{1}{2}E^2, p_t+ \frac{1}{2}E^2, p_t+ \frac{1}{2}E^2),
\end{equation}
where $\rho$ is the energy density, $p_r$ is the radial pressure,
$p_t$ is the tangential pressure and $E$ is the electric field
intensity. These quantities are measured relative to the comoving
fluid velocitry $u^i = e^{-\nu}\delta^i_0.$  For the line element
(\ref{eq:f1}) and matter distribution (\ref{eq:f2}) the Einstein
field equations can be expressed as
\begin{eqnarray}
\label{eq:f3} \frac{1}{r^{2}} \left[ r(1-e^{-2\lambda}) \right]' &
=&  \rho + \frac{1}{2}E^{2},\\
 \label{eq:f4} - \frac{1}{r^{2}} \left( 1-e^{-2\lambda} \right) +
\frac{2\nu'}{r}e^{-2\lambda} & = & p_r -\frac{1}{2}E^{2},\\
\label{eq:f5} e^{-2\lambda}\left( \nu'' + \nu'^{2} + \frac{\nu'}{r}-
\nu'\lambda' - \frac{\lambda'}{r} \right)
 & = & p_t + \frac{1}{2}E^{2}, \\
\label{eq:f6} \sigma & = & \frac{1}{r^{2}} e^{-\lambda}(r^{2}E)',
\end{eqnarray}
where primes denote differentiation with respect to $r$ and $\sigma$
is the  proper charge density. In the field equations
(\ref{eq:f3})-(\ref{eq:f6}), we are using units where the coupling
constant $\frac{8\pi G}{c^4}=1$ and the speed of light $c=1$.  The
system of equations (\ref{eq:f3})-(\ref{eq:f6}) governs the
behaviour of the gravitational field for an anisotropic charged
imperfect fluid. Note that the system (\ref{eq:f3})-(\ref{eq:f6})
becomes
\begin{eqnarray}
\label{eq:f7} \frac{1}{r^{2}} \left[ r(1-e^{-2\lambda}) \right]' &
=&  \rho ,\\
 \label{eq:f8} - \frac{1}{r^{2}} \left( 1-e^{-2\lambda} \right) +
\frac{2\nu'}{r}e^{-2\lambda} & = & p,\\
\label{eq:f9} e^{-2\lambda}\left( \nu'' + \nu'^{2} + \frac{\nu'}{r}-
\nu'\lambda' - \frac{\lambda'}{r} \right)
 & = & p,
\end{eqnarray}
for matter distributions with isotropic pressures $(p_r=p_t)$ in the
absence of charge $(E=0)$.

The mass contained within a radius $r$ of the sphere is defined as
\begin{equation}
\label{eq:f10} m(r)= \frac{1}{2}\int_0^r\omega^2 \rho(\omega)d\omega
.
\end{equation}
A different, but equivalent, form of the field equations is obtained
if we introduce a new independent variable $x$, and define functions
$y$ and $Z$, as follows
\begin{equation}
\label{eq:f11} x = Cr^2,~~ Z(x)  = e^{-2\lambda(r)} ~\mbox{and}~
A^{2}y^{2}(x) = e^{2\nu(r)},
\end{equation}
which was first suggested by Durgapal and Bannerji [28]. Then the
line element (\ref{eq:f1}) becomes
\begin{equation}
\label{eq:f12} ds^2 = -A^2 y^2 dt^2 + \frac{1}{4CxZ}dx^2 +
\frac{x}{C} (d\theta^2 +\sin^2\theta d\phi^2).
\end{equation}
In (\ref{eq:f11}) and (\ref{eq:f12}), the quantities $A$ and $C$ are
arbitrary constants. Under the transformation (\ref{eq:f11}), the
system (\ref{eq:f3})-(\ref{eq:f6}) becomes
\begin{eqnarray}
\label{eq:f13}\frac{1-Z}{x} - 2\dot{Z} & = & \frac{\rho}{C} +
 \frac{E^{2}}{2C}, \\
\label{eq:f14} 4Z\frac{\dot{y}}{y} + \frac{Z-1}{x} & = &
\frac{p_r}{C}
-  \frac{E^{2}}{2C}, \\
\label{eq:f15} 4x Z \frac{\ddot{y}}{y} +(4 Z+ 2 x \dot{Z})
\frac{\dot{y}}{y} + \dot{Z} &=&\frac{p_t}{C}+ \frac{E^2}{2C}, \\
\label{eq:f16} \frac{\sigma^{2}}{C} & = & \frac{4Z}{x} \left(x
\dot{E} + E \right)^{2},
\end{eqnarray}
where dots denote differentiation with respect to the variable $x$.
The mass function (\ref{eq:f10}) becomes
\begin{equation}
\label{eq:f17}m(x)=\frac{1}{4 C^{3/2}} \int_0^x\sqrt{w}\rho(w)dw,
\end{equation}
in terms of the new variables in (\ref{eq:f11}).

For a physically realistic relativistic star we expect that the
matter distribution should satisfy a barotropic equation of state
$p_r=p_r(\rho)$. For our purposes we assume  the linear equation of
state
\begin{equation}
\label{eq:f18} p_r = \alpha \rho - \beta ,
\end{equation}
where $\alpha$ and $\beta$ are constants. Then it is possible to
write the system (\ref{eq:f13})-(\ref{eq:f16}) in the simpler form
\begin{eqnarray}
\label{eq:f19}\frac{\rho}{C}&=& \frac{1-Z}{x}-2
\dot{Z}-\frac{E^2}{2C},\\
\label{eq:f20} p_r & =& \alpha \rho -\beta ,\\
\label{eq:f21}p_t &=& p_r +\Delta ,\\
 \Delta &=& 4 C x Z \frac{\ddot{y}}{y} + 2 C \left[x
\dot{Z}+ \frac{4Z}{(1+\alpha)}\right]\frac{\dot{y}}{y}\nonumber\\
\label{eq:f22}&& + \frac{(1+5 \alpha)}{(1+\alpha)}C\dot{Z}- \frac{C
(1-Z)}{x}+\frac{2 \beta}{(1+\alpha)},\\
\label{eq:f23}\frac{E^2}{2C}&=& \frac{1-Z}{x}
-\frac{1}{(1+\alpha)}\left[2 \alpha
\dot{Z}+ 4 Z \frac{\dot{y}}{y} +\frac{\beta}{C}\right],\\
\label{eq:f24} \frac{\sigma^2}{C}&=&4 \frac{Z}{x}(x\dot{E}+E)^2,
\end{eqnarray}
where the quantity $\Delta = p_t -p_r$ is the measure of anisotropy
in this model. In the system (\ref{eq:f19})-(\ref{eq:f24}), there
are eight independent variables  $(\rho, p_r, p_t, \Delta, E,
\sigma,y,Z)$ and only six independent equations. This suggests that
it is possible to specify two of the quantities involved in the
integration process. The resultant system will remain highly
nonlinear but it may be possible to generate exact solutions.

\section{Generating exact models}
We must make  physically reasonable choices for any two of the
independent variables and then solve the system
(\ref{eq:f19})-(\ref{eq:f24}) to generate exact models. In this
paper, we choose forms for the gravitational potential $Z$ and
electric field intensity $E$. We make the specific choices
\begin{eqnarray}
\label{eq:f25} Z&=&\frac{1+ (a-b)x}{1+
ax},\\
\label{eq:f26}\frac{E^2}{C}&=&\frac{k(3+ax)}{(1+ax)^2},
\end{eqnarray}
where $a, b$ and $k$ are real constants. The gravitational potential
$Z$ is regular at the origin and well behaved in the stellar
interior for a wide range of values for the parameters $a$ and $b$.
The electric field intensity  is continuous, bounded and a
decreasing function from the origin to the boundary of the sphere.
Therefore the forms chosen in (\ref{eq:f25})-(\ref{eq:f26}) are
physically reasonable. On substituting (\ref{eq:f25}) and
(\ref{eq:f26}) in (\ref{eq:f23}) we obtain
\begin{eqnarray}
 \frac{\dot{y}}{y}&=&
\frac{(1+\alpha)b}{4\left[1+(a-b)x\right]}+\frac{\alpha b}{2
(1+ax)\left[1+(a-b)x\right]}\nonumber\\
\label{eq:f27}&&-\frac{\beta
(1+ax)}{4C\left[1+(a-b)x\right]}-\frac{(1+\alpha)k
(3+ax)}{8(1+ax)\left[1+(a-b)x\right]},
\end{eqnarray}
which is a linear equation in the gravitational potential $y$. For
the integration  of  equation (\ref{eq:f27}) it is convenient to
consider three cases: $b=0, a=b$ and $a\neq b$.

\subsection{The case $b=0$}
When $b=0$,  (\ref{eq:f27}) becomes
\begin{equation}
\label{eq:f28} \frac{\dot{y}}{y}= -\frac{\beta
}{4C}-\frac{(1+\alpha)k (3+ax)}{8(1+ax)^2}
\end{equation}
with solution
\begin{equation}
\label{eq:f29}y=D
(1+ax)^{\frac{-k(1+\alpha)}{a}}\exp\left[\frac{2k(1+\alpha)}{a(1+ax)}-\frac{\beta
x}{4C}\right],
\end{equation}
where $D$ is  the constant of integration. We observe that $\rho =-
\frac{E^2}{2}$ for this case which we do not consider further to
avoid negative energy densities.

\subsection{The case $a=b$}
When $a=b$, (\ref{eq:f27}) becomes
\begin{equation}
\label{eq:f30} \frac{\dot{y}}{y}= \frac{(1+\alpha)a}{4}+\frac{\alpha
a}{2 (1+ax)}-\frac{\beta (1+ax)}{4C}-\frac{(1+\alpha)k
(3+ax)}{8(1+ax)}.
\end{equation}
On integrating (\ref{eq:f30}) we get
\begin{equation}
\label{eq:f31}y= D (1+ax)^{\frac{2 a
\alpha-(1+\alpha)k}{4a}}\exp[F(x)],
\end{equation}
where
\[F(x)= \frac{x}{8C}\left[-kC (1+\alpha)-2 \beta
+a(2C(1+\alpha)-\beta x)\right]\] and $D$ is the constant of
integration. Then we can generate an exact model for the system
(\ref{eq:f19})-(\ref{eq:f24}) as follows
\begin{eqnarray}
\label{eq:f32} e^{2\lambda} &=& 1+a x,\\
\label{eq:f33} e^{2\nu}&=& A^2 D^2 (1+ax)^{\frac{2a\alpha -
k(1+\alpha)}{2a}}\exp[2 F(x)],\\
\label{eq:f34} \frac{\rho}{C} &=& \frac{(2a-k)}{2}\frac{(3+a
x)}{(1+ax)^2},\\
\label{eq:f35} p_r &=& \alpha \rho -\beta ,\\
\label{eq:f36}p_t &=& p_r +\Delta ,\\
 \Delta &=& \frac{1}{16 C (1+ax)^3} \left\{ C^2 \left[
k^2 (1+\alpha)^2 x (3+ax)^2 \right. \right. \nonumber\\
&& \left. \left.+ 4a^2 x(3-8\alpha +9{\alpha}^2 + a^2
(1+\alpha)^2x^2 +2ax (2+3\alpha+3{\alpha}^2) )\right. \right.
\nonumber\\
&& \left. \left. -4k(12+a^3(1+\alpha)^2 x^3 +a^2x^2 (7+9\alpha +6
{\alpha}^2) +ax (12 +5 \alpha +9 {\alpha}^2)) \right]
\right.\nonumber\\
&&\left. -4Cx (1+ax)^2 [ (1+\alpha)(2a^2x -3k)- a \beta
(k(1+\alpha)-6
\alpha-4)]\right. \nonumber\\
\label{eq:f37}&& \left.+4 {\beta}^2 x(1+ax)^4\right\},\\
\label{eq:f38}\frac{E^2}{C} &=& \frac{k(3+ax)}{(1+ax)^2},
\end{eqnarray}
in terms of elementary functions.

The solution (\ref{eq:f32})-(\ref{eq:f38}) may be used to model a
charged anisotropic star with a linear equation of state.
 In this case the mass function  is
\begin{equation}
\label{eq:f39}m(x)=\frac{(2a-k)x^{3/2}}{4C^{3/2}(1+ax)},
\end{equation}
which is similar to forms used in other investigations. The
gravitational potentials and matter variables are continuous and
well behaved in the stellar interior. Note that when $k=0$ the model
(\ref{eq:f32})-(\ref{eq:f38})  reduces to a solution for uncharged
anisotropic  stars. Equation (\ref{eq:f37}) yields
\begin{eqnarray}
\Delta &=& \frac{1}{4 C (1+ax)^3} \left\{ C^2 a^2 x \left[3-8\alpha
+9{\alpha}^2 + a^2 (1+\alpha)^2x^2 +2ax (2+3\alpha+3{\alpha}^2)
\right]
\right.\nonumber\\
\label{eq:f40} &&\left. -2Cx (1+ax)^2 [ (1+\alpha)a^2x + a \beta (3
\alpha+2)] + {\beta}^2 x(1+ax)^4\right\}
\end{eqnarray}
when $k=0$ so that the model is necessarily anisotropic with $\Delta
\neq 0$ in general even in the simpler case of uncharged matter.
Some treatments of the physical properties of anisotropic spheres in
general relativity include the investigations of Dev and Gleiser
[29, 30], Mak and Harko [31, 32], Chaisi and Maharaj [33, 34] and
Maharaj and Chaisi [35] with $\Delta \neq 0$.

\subsection{The case $a\neq b$}
On integrating (\ref{eq:f27}) we get
\begin{equation}
\label{eq:f41}y=D (1+ax)^m [1+(a-b)x]^n \exp\left[\frac{-a\beta
x}{4C (a-b)}\right]
\end{equation}
where $D$ is the constant of integration, and $m$ and $n$ are given
by
\begin{eqnarray}
m &=& \frac{2 \alpha b -(1+\alpha)k}{4b},\nonumber\\
n&=& \frac{1}{8b C(a-b)^2} \left[2a^2 C (k(1+\alpha)-2\alpha b) -
a b C (5k(1+\alpha) -2b (1+5\alpha))\right.\nonumber\\
&& \left.+b^2 (3kC(1+\alpha) - 2bC
(1+3\alpha)+2\beta)\right].\nonumber
\end{eqnarray}
Then we can generate an exact model for the system
(\ref{eq:f19})-(\ref{eq:f24}) in the form
\begin{eqnarray}
\label{eq:f42} e^{2\lambda} &=& \frac{1+ax}{1+(a-b)x},\\
\label{eq:f43} e^{2\nu}&=& A^2 D^2 (1+ax)^{2m} [1+(a-b)x]^{2n}
\exp{\left[\frac{-a\beta x}{2C(a-b)}\right]},\\
\label{eq:f44} \frac{\rho}{C} &=& \frac{(2b-k)}{2}\frac{(3+a
x)}{(1+ax)^2},\\
\label{eq:f45} p_r &=& \alpha \rho -\beta ,\\
\label{eq:f46}p_t &=& p_r +\Delta ,\\
 \label{eq:f47}\Delta &=& \frac{-bC}{(1+ax)}
 -\frac{bC(1+5\alpha)}{(1+\alpha)(1+ax)^2}+\frac{2\beta}{(1+\alpha)}\nonumber\\
 && +
 \frac{Cx[1+(a-b)x]}{(1+ax)}\left[4\left(\frac{a^2m(m-1)}{(1+ax)^2} +\frac{2a(a-b)mn}{(1+ax)[1+(a-b)x]}
 \right. \right. \nonumber\\
 &&\left. \left.+\frac{(a-b)^2n(n-1)}{[1+(a-b)x]^2} \right)
 - \frac{2a\beta (a(m+n)[1+(a-b)x]-bn)}{(a-b)C(1+ax)[1+(a-b)x]}
 +\frac{a^2{\beta}^2}{4C^2 (a-b)^2}\right]\nonumber\\
&& -\frac{4 [1+ax(2+(a-b)x)]-b(5+
\alpha)x}{2(a-b)(1+\alpha)(1+ax)^3[1+(a-b)x]}\times \nonumber\\
&&\left[-4b^2Cn +a^3 x(-4C(m+n)+\beta x)+a^2 (4C(m+n)(2bx-1)+\beta
(2-bx)x)\right. \nonumber\\
&& \left. + a (-4b^2C(m+n)x+ \beta +b(4Cm+8Cn- \beta x))\right],\\
\label{eq:f48}\frac{E^2}{C} &=& \frac{k(3+ax)}{(1+ax)^2},
\end{eqnarray}
in terms of elementary functions.

Therefore we have generated a second class of solutions
(\ref{eq:f42})-(\ref{eq:f48}) that models a charged anisotropic star
with a linear equation of state. The mass function has the form
\begin{equation}
\label{eq:f49} m(x)=\frac{(2b-k)x^{3/2}}{4C^{3/2}(1+ax)}.
\end{equation}
The form of the mass function (\ref{eq:f49}) represents an energy
density which is monotonically decreasing in the stellar interior
and remains finite at the centre $x=0$. It is physically reasonable
and has been used in the past to study the properties of isotropic
fluid spheres: Matese and Whitman [36] generated equilibrium
configurations in general relativity, Finch and Skea [37] studied
neutron star models and Mak and Harko [32] analysed anisotropic
relativistic stars with this form of mass function. Lobo [10]
demonstrated that (\ref{eq:f49}) is consistent with stable dark
energy stars which generalises the gravastar  model of Mazur and
Mottola [38]. It was then shown that large stability regions exist
close to the event horizon thereby making it difficult to
distinguish dark energy stars from black holes. Sharma and Maharaj
[11] found a new class of exact solutions to Einstein equations that
can be applied to strange stars with quark matter with this mass
distribution. Consequently the mass function (\ref{eq:f49}) is of
astrophysical importance in the description of compact objects.

It is interesting to observe that for particular parameter values we
can regain uncharged anisotropic and isotropic models $(k=0)$ from
our general solution (\ref{eq:f42})-(\ref{eq:f48}).  We regain the
following particular cases of physical interest:

\subsubsection{Sharma and Maharaj model}
If we set $\beta =\alpha {\rho}_s$ then
\[p_r= \alpha (\rho -{\rho}_s),\]
where ${\rho}_s$ is the density at the surface $r=s$. Thus we regain
the equation of state of Sharma and Maharaj [11]. Then by setting
$C=1$ and $A^2D^2=B$ we find that the line element is of the form
\begin{eqnarray} ds^2 &=&-B (1+a r^2)^{\alpha} [1+
(a-b)r^2]^\gamma \exp\left(\frac{- a \beta r^2}{2(a-b)}\right)
dt^2\nonumber\\
\label{eq:f50}&&+ \frac{1+ax}{1+(a-b)x}dr^2+
 r^{2}(d\theta^{2} + \sin^{2}{\theta} d\phi^{2}),
 \end{eqnarray}
 where
 \[\gamma =\frac{5ab \alpha -2 a^2\alpha -3b^2 \alpha +ab -b^2 +b
 \beta}{2(a-b)^2}.\]
 The line element (\ref{eq:f50}) corresponds to
 the uncharged anisotropic model of Sharma and Maharaj [11]. They
 showed that this solution may be used to describe compact objects
 such as strange stars with a linear equation of state with quark
 matter.

\subsubsection{Lobo model}
If we set $\beta =0$ then
\[p_r=\alpha \rho\]
and we regain the equation of state studied by Lobo [10]. Then on
setting  $a=2b, C=1$ and $A^2D^2=1$ we generate the metric
\begin{equation}
\label{eq:f51}ds^2= -(1+br^2)^{(1- \alpha)/2}(1+2br^2)^{\alpha}dt^2
+\left(\frac{1+2br^2}{1+br^2}\right)dr^2+
 r^{2}(d\theta^{2} + \sin^{2}{\theta} d\phi^{2}).
 \end{equation}
The line element (\ref{eq:f51}) corresponds to the uncharged
anisotropic model of Lobo [10]. We point out that the line element
(\ref{eq:f51}) serves as an interior solution with $\alpha < -
\frac{1}{3}$ which may be matched to an exterior Schwarzschild
solution in a model for dark energy stars. Lobo [10] proved that
stability regions exist for dark energy stars by selecting
particular values of $\alpha$ in a graphical analysis.

\subsubsection{Isotropic models}
In general $\Delta \neq 0$ so that the model remains anisotropic.
However for particular parameter values we can show that $\Delta =0$
in the relevant limit in the general solution
(\ref{eq:f42})-(\ref{eq:f48}). If we set $a=0$ and $b=1$ then we
obtain
\begin{eqnarray}
m&=&\frac{\alpha}{2}\nonumber\\
n&=& \frac{1}{4C}[\beta -(1+3 \alpha)C] \nonumber\\
\label{eq:f52}\Delta &=& \frac{x}{4C(1-x)}[\beta
-3(1+\alpha)C][\beta -(1+3 \alpha)C].
\end{eqnarray}
Two different cases arise as a consequence of (\ref{eq:f52}) if we
set $\Delta =0$.

In the first case we observe that when $\beta =0$ and $\alpha =-1$
then $\Delta =0$. The  equation of state becomes $p_r (=p_t)=-\rho$.
In this case the line element becomes
\begin{equation}
\label{eq:f53}ds^2 = -\left(1-\frac{r^2}{R^2}\right)dt^2
+\left(1-\frac{r^2}{R^2}\right)^{-1}dr^2+
 r^{2}(d\theta^{2} + \sin^{2}{\theta} d\phi^{2}),
 \end{equation}
where we have set $A=D=1$ and $C=\frac{1}{R^2}$. The metric
(\ref{eq:f53}) corresponds to the familiar isotropic uncharged  de
Sitter model.

In the second case we see that when $\beta =0$ and $\alpha
=-\frac{1}{3}$ then $\Delta =0$. The equation of state becomes $p_r
(=p_t)=-\frac{1}{3}\rho$. In this case the line element becomes
\begin{equation}
\label{eq:f54}ds^2 = -A^2dt^2
+\left(1-\frac{r^2}{R^2}\right)^{-1}dr^2+
 r^{2}(d\theta^{2} + \sin^{2}{\theta} d\phi^{2}),
 \end{equation}
where we have set $D=1$ and $C=\frac{1}{R^2}$. The metric
(\ref{eq:f54}) corresponds to the well known isotropic uncharged
Einstein model.

\section{Physical analysis}
The solutions found in this paper may be connected to the
Einstein-Maxwell equations for the exterior of our source. We need
to match the Reissner-Nordstrom exterior spacetime
\[ds^2=-\left(1-\frac{2M}{r}+\frac{Q^2}{r^2}\right)dt^2
+\left(1-\frac{2M}{r}+\frac{Q^2}{r^2}\right)^{-1}dr^2
+r^2\left(d\theta^{2} + \sin^{2}{\theta} d\phi^{2}\right)\] to the
interior spacetime (\ref{eq:f1}) across the boundary $r=R$. This
generates the conditions
\begin{eqnarray}
1-\frac{2M}{R}+\frac{Q^2}{R^2}&=& A^2y^2(CR^2) \nonumber\\
\left(1-\frac{2M}{R}+\frac{Q^2}{R^2}\right)^{-1}&=&\frac{1+aCR^2}{1+(a-b)CR^2}
\nonumber
\end{eqnarray}
which relates the constants $a, b, A, C, D, \alpha$ and $\beta$.
This demonstrates that the continuity of the metric coefficients
across the boundary of the star $r=R$ is easily satisfied as there
are sufficient number of free parameters. If there is a surface
layer of charge then the pressure may be nonzero which would place
restrictions on the function $\nu$ through the matching conditions
at the boundary. However the number of free parameters available
easily satisfies the necessary conditions that arise for a
particular model under investigation.

We now  briefly consider the physical behaviour of the models
generated in Section 3 for the case $a \neq b$. From the explicit
forms (\ref{eq:f42}) and (\ref{eq:f43}) we can easily see that the
gravitational potentials $e^{2\nu}$ and $e^{2\lambda}$ are
continuous, well behaved and nonsingular at the origin. The energy
density $\rho$ is continuous and monotonically decreasing from the
centre to the boundary of the star, which is a necessary condition
for a realistic model. The radial pressure $p_r$ also has the same
feature because $\rho$ and $p_r$ are linked by a linear equation of
state. The tangential pressure $p_t$ is also nonsingular at the
origin and continuous for a wide range of the parameters $a, b$ and
$k$. To maintain the usual casuality condition we must place the
restriction  that $0 \leq \alpha \leq 1$ if we require
$\frac{dp_r}{d\rho}\leq 1$.  However note that our models do allow
for $\alpha < 0$ in the case  of anisotropic dark energy stars. The
form chosen for electric field intensity $E$ is physically
reasonable and describes a decreasing function .

With the help of  a particular example we can demonstrate the above
features graphically. Figures 1-4 represent the energy density, the
radial pressure, the tangential pressure and the electric field
intensity, respectively.  To plot the graphs   we choose the
parameters $a=3, b=2.15, \alpha = 0.33, \beta = \alpha {\rho}_s
=0.198, C=1$ and $k=0.2$, where ${\rho}_s$ is the density at the
boundary $r=s=1.157$. Note that our choice of $\alpha =0.33$ ensures
that both the radial pressure and the tangential pressure for the
neutral sphere vanish at the boundary. We observe  from Figures 1-4
that the matter variables $\rho, p_r, p_t$ and $E$ have the
appropriate features to describe a compact relativistic sphere.
Solid lines represent uncharged matter and dashed lines include the
effect of charge in Figures 1-3. We observe that the effect of $E$
is to produce lower values for $\rho, p_r$ and $p_t$ when compared
to the case of uncharged matter. In Figure 5 we have plotted the
measure of anisotropy $\Delta$ for the same parameter values used
above. Note that the effect of the  electromagnetic field is to
increase the magnitude of $\Delta$ which affects the behaviour of
$p_t$.

\vspace{1cm}
\begin{figure}[thb]
\vspace{1.5in} \includegraphics{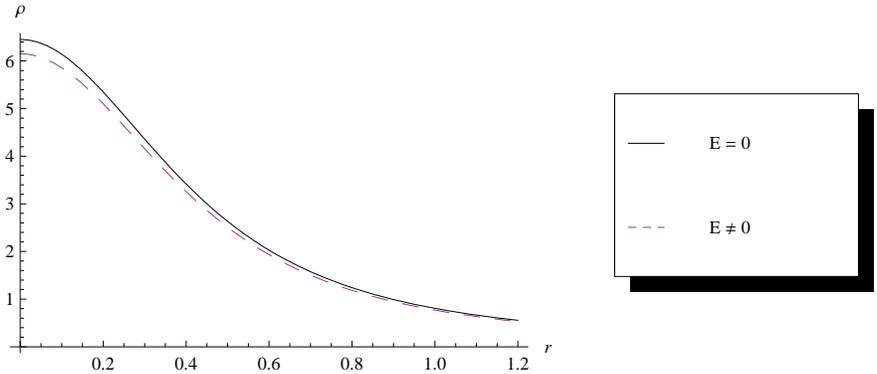}  \vspace{2mm}
\caption{\label{Gr-density} Energy density.}
\end{figure}

\vspace{1cm}
\begin{figure}[thb]
\vspace{1.5in} \includegraphics{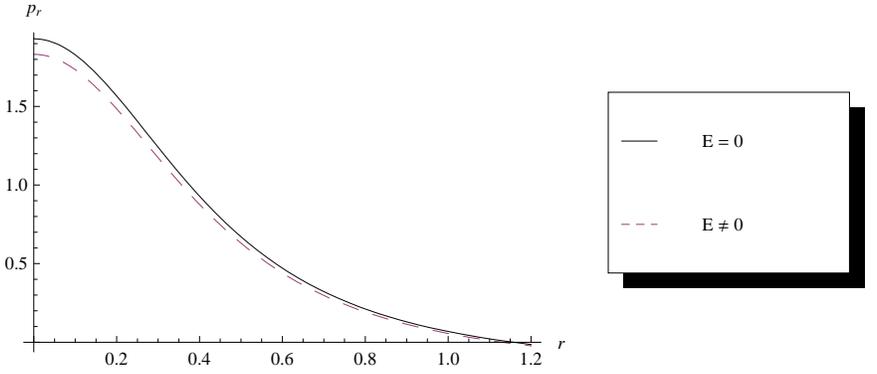}  \vspace{2mm}
\caption{\label{Gr-pressurer} Radial pressure.}
\end{figure}

\vspace{1cm}
\begin{figure}[thb]
\vspace{1.5in} \includegraphics{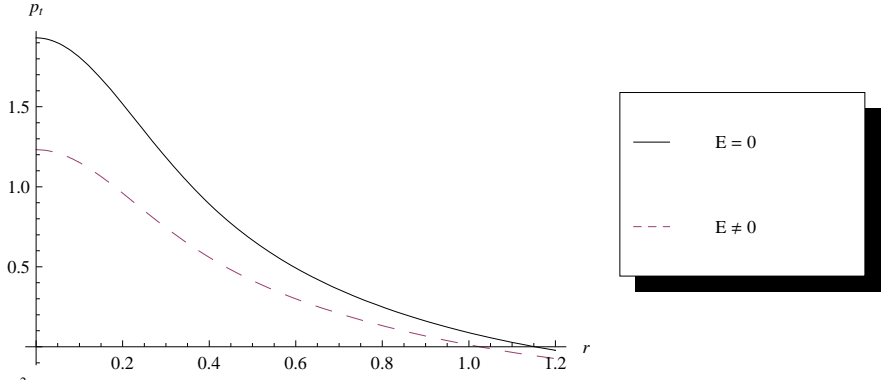}  \vspace{2mm}
\caption{\label{Gr-del-2} Tangential pressure.}
\end{figure}

\vspace{1cm}
\begin{figure}[thb]
\vspace{1.5in} \includegraphics{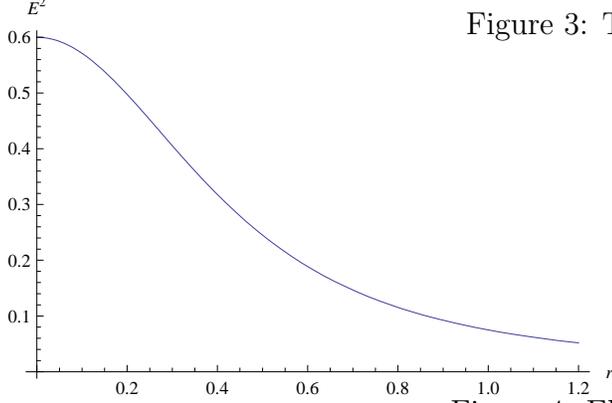}  \vspace{2mm}
\caption{\label{Gr-Electricfield} Electric field intensity.}
\end{figure}

\vspace{1cm}
\begin{figure}[thb]
\vspace{1.5in} \includegraphics{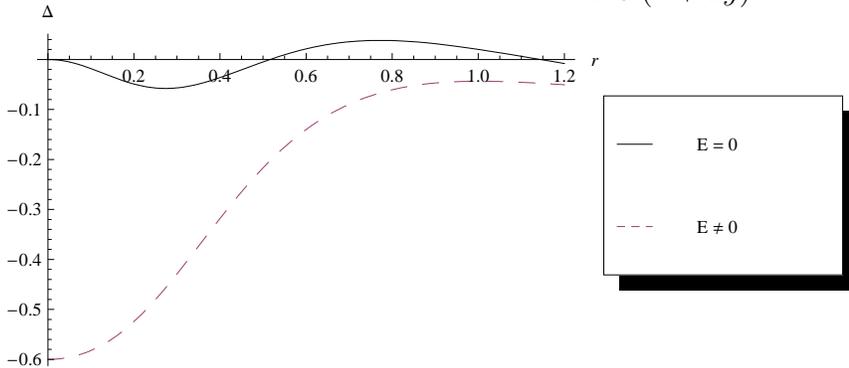}  \vspace{2mm} \caption{\label{Gr-del-2}
Measure of anisotropy.}
\end{figure}

\newpage
\section{Stellar structure}
In this section we show that the solutions generated in this paper
can be used to describe realistic compact objects. In particular we
seek to compare our results with those of Sharma and Maharaj [11]
since they regain values for the stellar mass agreeable with
observations. To achieve consistency with Sharma and Maharaj [11] we
introduce the transformations
\[\tilde{a}=a R^2,~
\tilde{b}=bR^2,~ \tilde{\beta}=\beta R^2,~\tilde{k}=k R^2.\] Under
these transformations the energy density becomes
\begin{equation}
\label{eq:f55}\rho =\frac{(2 \tilde{b}-\tilde{k})(3+\tilde{a}y)}{2
R^2 (1+\tilde{a}y)^2},
\end{equation}
and the mass contained within a radius $s$ has the form
\begin{equation}
\label{eq:f56}M=\frac{(2\tilde{b}
-\tilde{k})s^3/R^2}{4(1+\tilde{a}s^2/R^2)},
\end{equation}
where we have set $C=1$ and $y=\frac{r^2}{R^2}$. When $\tilde{k}=0
~(\mbox{or}~E=0)$, (\ref{eq:f55}) and (\ref{eq:f56}) reduce to the
expressions of Sharma and Maharaj [11]:
\begin{equation}
\rho =
\frac{\tilde{b}(3+\tilde{a}y)}{R^2(1+\tilde{a}y)^2},~~M=\frac{\tilde{b}s^3/R^2}{2(1+\tilde{a}s^2/R^2)}
\end{equation}
which gives the density $\rho$ and mass $M$ of an uncharged star of
radius $s$.

If we choose $\tilde{a}=53.34, \tilde{b}=54.34,R=43.245$ km
 and $s=7.07$ km then we can produce an uncharged model
 $(\tilde{k}=0)$ with mass $M= 1.433M_{\bigodot}$ and central
 density $\rho_c =4.672\times 10^{15}$gcm$^{-3}$. The
 corresponding value of $\alpha =0.437$ is obtained by requiring
 that the anisotropy vanishes at the boundary.  To simplify
 comparison with Sharma and Maharaj [11] we have used the same values of
 $\tilde{a}, \tilde{b}, R$ and $s$; however our value for $\alpha$
 is a correction. It should be noted that these results are
 consistent with the equation of state for strange matter formulated
 by Dey $et~al$ [39]. This has astrophysical significance as
 their model has been used to describe the X-ray binary pulsar SAX
 J1808.4-3658. When the charge is nonzero we set $\tilde{k}=37.403$
 and then we obtain the mass $M= 0.940M_{\bigodot}$ and central
 density $\rho_c =3.064\times 10^{15}$gcm$^{-3}$. The values
 for $M$ and $\rho_c$ generalise the figures of Sharma and Maharaj
 [11] to include the effect of the electromagnetic field. Choosing
 different set of values for the parameters will produce different
 results as shown in Table 1.  Note that the values presented in
 Table 1 correspond to a star of radius $s=7.07$km. The value of $\tilde{k}= 37.403$
is selected, in generating Table 1, so that the density and mass of
the Sharma and Maharaj [11] analysis is regained for uncharged
matter. Furthermore, the value of $\tilde{k}= 37.403$ with $E=0$
generates a star of mass $1.433 M_{\bigodot}$ which is the same as
the strange star model of Dey $et~al$ [39]. With this value of
$\tilde{k}$ we find that the star has mass $0.940 M_{\bigodot}$ in
the presence of charge so that the stellar core has a lower density
which represents a weaker field. This is consistent as the effect of
the electromagnetic field is repulsive.

We observe that the values for the mass in the presence of charge
$(E\neq 0)$ is always less than the uncharged case. The central
density of the charged sphere is also less than the uncharged case.
Sharma and Maharaj [11] showed that anisotropy affects the mass and
central densities of massive objects. We have shown that the
inclusion of the electromagnetic field also affects $M$ and
$\rho_c$. Both anisotropy and charge are physical quantities that
affect the range of degenerate states in our model. For the
calculation of mass and central density we  have set $s=7.07$ km,
$R=43.245$ km, $\tilde{k}=37.403$ and ${\rho}_s= 1.17119 \times
10^{15}\mbox{gcm}^{-3}$ for the uncharged case.

\begin{table}
\begin{tabular}{|c|c|c|c|c|c|c|}
\hline  $\tilde{b}$& $\tilde{a}$ & $\alpha$ & ${\rho}_c$ & $M$
& ${\rho}_c$ & $M$ \\
 & & & $(\times 10^{15}~\mbox{gcm}^{-3})$ &
($M_{\bigodot}$)& $
(\times 10^{15}~\mbox{gcm}^{-3})$ & ($M_{\bigodot}$)\\
& & & $E=0$ & $E=0$ & $E\neq 0$ & $E\neq 0$\\
\hline
30 ~~~~&    23.681 & 0.401 & 2.579 & 1.175 & 0.971 & 0.443\\
40 ~~~~&    36.346 & 0.400 & 3.439 & 1.298 & 1.831 & 0.691\\
50 ~~~~&    48.307 & 0.424 & 4.298 & 1.396 & 2.691 & 0.874\\
54.34  &    53.340 & 0.437 & 4.671 & 1.433 & 3.064 & 0.940\\
60 ~~~~&    59.788 & 0.457 & 5.158 & 1.477 & 3.550 & 1.017\\
70~~~~ &    70.920 & 0.495 & 6.017 & 1.546 & 4.410 & 1.133\\
80 ~~~~&    81.786 & 0.537 & 6.877 & 1.606 & 5.269 & 1.231\\
90 ~~~~&    92.442 & 0.581 & 7.737 & 1.659 & 6.129 & 1.314\\
100~~~~ &   102.929 & 0.627 & 8.596 & 1.705 & 6.989 & 1.386\\
183 ~~~~&   186.163 & 1.083 &15.730 & 1.959 &14.124 & 1.759 \\
\hline
\end{tabular}
\centering \caption{Central density and mass for different
anisotropic stellar models for neutral and charged bodies}
\end{table}

\section{Conclusion}
We have found a general framework for the Einstein-Maxwell system of
equations with a linear equation of state for anisotropic matter
distributions in the presence of the electromagnetic field. Three
new classes of exact solutions have been generated to this system of
nonlinear equations. We have shown that these classes of solutions
satisfy the necessary physical requirements in the  description of a
charged compact objects with anisotropic matter distribution. We
have demonstrated that our models yield stellar structures with
masses and densities  consistent with the Dey $et~ al$ [38] and
Sharma and Maharaj [11] models in the limit of vanishing charge.
Therefore it is likely that our solutions may be helpful in the
gravitational description of stellar bodies such as SAX
J1808.4-3658. The solutions obtained may be useful to model the
interior of charged relativistic quark stars with anisotropic matter
distribution. Our models contain the uncharged anisotropy models of
Sharma and Maharaj [11] and Lobo [10] which describe quark stars and
strange matter stars. We believe that the general class of exact
solutions found in this paper may assist in more detailed studies of
relativistic stellar bodies and allows for different matter
distributions because of the form of the linear equation of state
chosen. \\
~\\
{\bf Acknowledgements}\\
ST thanks the National Research Foundation and the University of
KwaZulu-Natal for financial support, and is grateful to Eastern
University, Sri Lanka for study leave. SDM acknowledges that this
work is based upon research supported by the South African Research
Chair Initiative of the Department of Science and
Technology and the National Research Foundation.\\
~\\


\thebibliography{}
\bibitem{1} Bowers R L and Liang E P T 1974 \emph{Astrophys. J} \textbf{188} 657
\bibitem{2} Ruderman R 1972 \emph{Ann. Rev. Astron. Astrophys.}
\textbf{10} 427
\bibitem{3} Kippenhahn R and Weigert A 1990 \emph{Stellar Structure and
Evolution} (Berlin: Springer Verlag)
\bibitem{4} Sokolov A I 1980 \emph{JETP} \textbf{79} 1137
\bibitem{5} Sawyer R F 1972 \emph{Phys. Rev. Lett.} \textbf{29}
382
\bibitem{6} Binney  J and Tremaine J S 1987 \emph{Galactic
Dynamics} (Princeton: Princeton University Press)
\bibitem{7} Herrera L and Santos N O 1995 \emph{Astrophys. J} \textbf{438}
308
\bibitem{8} Letelier P 1980 \emph{Phys. Rev. D} \textbf{22} 807
\bibitem{9} Weber F 1999 \emph{Pulsars as Astrophysical Observatories for Nuclear and Particle Physics}
(Bristol: IOP Publishing)
\bibitem{10} Lobo F S N 2006 \emph{Class. Quantum Grav.}
\textbf{23} 1525
\bibitem{11} Sharma R and Maharaj S D 2007 \emph{Mon. Not. R. Astron.
Soc.} \textbf{375} 1265
\bibitem{12} Ivanov B V 2002 \emph{Phys. Rev. D} \textbf{65} 104001
\bibitem{39} Thirukkanesh S and Maharaj S D 2006 \emph{Class. Quantum
Grav.} \textbf{23} 2697
\bibitem{13} Komathiraj K and Maharaj S D  2007 \emph{J. Math. Phys.} \textbf{48}
042501
\bibitem{14} Sharma R, Mukherjee S and Maharaj S D 2001 \emph{Gen. Relativ.
Gravit.} \textbf{33} 999
\bibitem{15} Patel L K and Koppar S K 1987 \emph{Aust. J. Phys.}
\textbf{40} 441
\bibitem{16} Tikekar R and Singh G P 1998 \emph{Grav. Cosmol.} \textbf{4} 294
\bibitem{17} Gupta Y K and Kumar M 2005 \emph{Gen. Relativ. Gravit.}
\textbf{37} 233
\bibitem{18} Sharma R, Karmakar S and  Mukherjee S  2006 \emph{Int. J. Mod. Phys. D} \textbf{15}
405
\bibitem{19} Sharma R and Mukherjee S  2001 \emph{Mod. Phys. Lett. A} \textbf{16}
1049
\bibitem{20} Sharma R and Mukherjee S  2002 \emph{Mod. Phys. Lett. A} \textbf{17}
2535
\bibitem{21} Thomas V O, Ratanpal B S and Vinodkumar P C  2005 \emph{Int. J. Mod. Phys. D} \textbf{14}
85
\bibitem{22} Tikekar R and Thomas V O  1998 \emph{Pramana - J. Phys.} \textbf{50}
95
\bibitem{23} Paul B C and  Tikekar R 2005 \emph{Grav. Cosmol.} \textbf{11} 244
\bibitem{24} Mak M K and Harko T   2004 \emph{Int. J. Mod. Phys. D} \textbf{13}
149
\bibitem{25} Komathiraj K and Maharaj S D  2007 \emph{Int. J. Mod. Phys. D} \textbf{16}
1803
\bibitem{26} Witten E  1984 \emph{Phys. Rev. D} \textbf{30}
272
\bibitem{27} Durgapal M C and Bannerji R 1983 \emph{Phys. Rev. D} \textbf{27}
328
\bibitem{28} Dev K and Gleiser M 2002 \emph{Gen. Relativ. Gravit.} \textbf{34}
1793
\bibitem{29} Dev K and Gleiser M 2003 \emph{Gen. Relativ. Gravit.} \textbf{35}
1435
\bibitem{30} Mak M K and Harko T 2002 \emph{Chin. J. Astron. Astrophys.} \textbf{2}
248
\bibitem{31} Mak M K and Harko T 2003 \emph{Proc. Roy. Soc. Lond. A} \textbf{459}
393
\bibitem{32} Chaisi M and Maharaj S D  2005 \emph{Gen. Relativ. Gravit.} \textbf{37}
1177
\bibitem{33} Chaisi M and Maharaj S D  2006 \emph{Pramana - J. Phys.} \textbf{66}
609
\bibitem{34} Maharaj S D and Chaisi M 2006 \emph{Gen. Relativ. Gravit.} \textbf{38}
1723
\bibitem{35} Matese J J and Whitman P G 1980  \emph{Phys. Rev. D} \textbf{22}
1270
\bibitem{36} Finch M R and Skea J E F 1989 \emph{Class. Quantum Grav.}
\textbf{6} 467
\bibitem{37} Mazur P O and Mottola E 2004 \emph{Preprint: gr-qc/0405111}
\bibitem{38} Dey M, Bombaci I, Ray S and Samanta B C 1998 \emph{Phys. Lett. B}
\textbf{438} 123; addendum: 1999, 447, 352; erratum: 1999, 467, 303
\end{document}